\documentclass[final]{aipproc}
\layoutstyle{6x9}
\begin{document}

\thispagestyle{plain}
\setcounter{page}{278}

\title{A Mott-like State of Molecules}



\author{S. D{\"u}rr}
{address={Max-Planck-Institut f{\"u}r Quantenoptik, Hans-Kopfermann-Stra{\ss}e 1, 85748 Garching, Germany}}
\author{T. Volz}{}
\author{N. Syassen}{}
\author{D.~M. Bauer}{}
\author{E. Hansis}{}
\author{G. Rempe}{}

\begin{abstract}
We prepare a quantum state where each site of an optical lattice is occupied by exactly one molecule. This is the same quantum state as in a Mott insulator of molecules in the limit of negligible tunneling. Unlike previous Mott insulators, our system consists of molecules which can collide inelastically. In the absence of the optical lattice these collisions would lead to fast loss of the molecules from the sample. To prepare the state, we start from a Mott insulator of atomic $^{87}$Rb with a central region, where each lattice site is occupied by exactly two atoms. We then associate molecules using a Feshbach resonance. Remaining atoms can be removed using blast light. Our method does not rely on the molecule-molecule interaction properties and is therefore applicable to many systems.
\end{abstract}

\maketitle

A variety of interesting proposals for quantum information processing and quantum simulations \cite{demille:02,goral:02,lee:05,micheli:06,barnett:06} require as a prerequisite a quantum state of ultracold polar molecules in an optical lattice, where each lattice site is occupied by exactly one molecule. A promising strategy for the creation of such molecules is based on association of ultracold atoms using a Feshbach resonance \cite{regal:03,herbig:03,duerr:04,strecker:03,cubizolles:03,jochim:03a,xu:03,koehler:06} or photoassociation and subsequent transfer to a much lower ro-vibrational level using Raman transitions \cite{sage:05}. If the molecule-molecule interactions are predominantly elastic and effectively repulsive, then a state with one molecule per lattice site can finally be obtained using a quantum phase transition from a superfluid to a Mott insulator by ramping up the depth of an optical lattice \cite{greiner:02}. However, many molecular species do not have such convenient interaction properties, so that alternative strategies are needed. Here, we demonstrate a technique that is independent of the molecule-molecule interaction properties. The technique relies on first forming an atomic Mott insulator and then associating molecules. Several previous experiments \cite{thalhammer:06,rom:04,stoferle:06,winkler:06,ryu:cond-mat/0508201-aipproc} associated molecules in an optical lattice, but none of them demonstrated the production of a quantum state with exactly one molecule per lattice site. Another interesting perspective of the state prepared here is that after Raman transitions to the ro-vibrational ground state, one could lower the lattice potential and obtain a Bose-Einstein condensate (BEC) of molecules in the ro-vibrational ground-state \cite{jaksch:02,damski:03}. Central results of our work were previously reported in Ref.~\cite{volz:06}.

The experiment begins with the creation of a BEC of $^{87}$Rb atoms in a magnetic trap \cite{marte:02}. Once created, the BEC is transferred into a crossed-beam optical dipole trap and then a magnetic field of approximately 1000~G is applied. The atoms are transferred \cite{volz:03} to the absolute ground state $|F=1,m_F=1\rangle$, which has a Feshbach resonance at 1007.4~G \cite{marte:02} with a width of 0.2~G \cite{volz:03,duerr:04a}. Next, a simple-cubic optical lattice is created by illuminating the BEC with three retro-reflected light beams at a wavelength of $2\pi/k=830$~nm. The behavior of bosons in this optical lattice is described by the Bose-Hubbard Hamiltonian \cite{jaksch:98}. The relevant parameters for atoms in the lattice are the lattice depth $V_0$ \cite{greiner:02}, the amplitude $J$ for tunneling between neighboring lattice sites, and the on-site interaction matrix element $U$. Note that the polarizability of a Feshbach molecule is approximately twice as large as for one atom. Hence, a molecule experiences a lattice depth of $2V_0$. Combined with the fact that a molecule has twice the atomic mass, this drastically reduces the tunneling amplitude for molecules compared to atoms, e.g., at a typical lattice depth of $V_0=24 E_r$, the tunneling amplitude for an atom is $J=2\pi\hbar\times 4$~Hz, whereas the tunneling amplitude for a molecule at the same intensity of the lattice light is $J_m=2\pi\hbar \times 0.3$~mHz. Here, $E_r=\hbar^2k^2/(2m)$ is the recoil energy and $m$ is the mass of one atom. We experimentally calibrate the lattice depth with the method shown in Fig.~3 of Ref.~\cite{hecker:02}.

\begin{figure}[h]
\includegraphics[height=3cm]{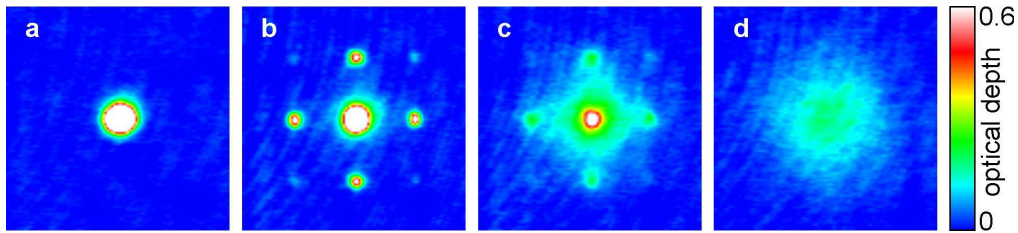}
\caption{\label{fig-atomic-MI}
Atomic Mott insulator. The lattice depth is slowly ramped up and then suddenly switched off. After some time of flight, an absorption image is taken. Parts (a)-(d) show images corresponding to a maximum lattice depth of $V_0/E_r=0$, 5, 12, and 22. For small lattice depth, the system is superfluid and diffraction from the lattice light leads to an atomic interference pattern. For a deep lattice, however, the system is in the Mott phase, corresponding to a fixed atom number per lattice site. This results in a maximum uncertainty of the relative phase between sites and therefore the interference pattern is washed out.
 }
\end{figure}

We create an atomic Mott insulator at a magnetic field of 1008.8~G by slowly (within 80~ms) ramping up the depth of the optical lattice. We carefully checked that a Mott insulator is obtained by repeating all measurements presented in Ref.~\cite{greiner:02}. In particular, we used a time-of-flight method to measure the momentum distribution after suddenly switching off the optical lattice potential, as seen in Fig.~\ref{fig-atomic-MI}.

\thispagestyle{plain}

The system has an external harmonic confinement due to the finite waist of the lattice beams and due to the additional dipole trap. This makes the atomic Mott insulator inhomogeneous. It consists of shells of constant lattice filling with exactly $n$ particles per lattice site. Neighboring shells are connected by narrow superfluid regions. By choosing appropriate parameters \cite{volz:06}, we ensure that the core of the cloud contains exactly $n=2$ atoms per lattice site. This core contains 47(3)\% out of a total population of $10^5$ atoms.

After preparing the atomic Mott insulator at a magnetic field of $B=1008.8$~G, molecules are associated as described in Ref.~\cite{duerr:04}. To this end, the magnetic field is slowly (at 2~G/ms) ramped across the Feshbach resonance at 1007.4~G to a final value of $B=1006.6$~G. At lattice sites with a filling of $n=1$, this has no effect. At sites with $n>1$, atom pairs are associated to molecules. If the site contained $n>2$ atoms, then the molecule can collide with other atoms or molecules at the same lattice site. As the molecules are associated in a highly-excited ro-vibrational state, the collisions are likely to be inelastic. This leads to fast loss of the molecule and its collision partner from the trap. The association ramp lasts long enough to essentially empty all sites with $n>2$ atoms. For lattice sites with $n=2$ atoms, the association efficiency is close to unity. At a lattice depth of $V_0=24E_r$ for atoms, the tunneling amplitude for molecules (see above) is negligible compared to the hold time between association and dissociation, so that the positions of the molecules are frozen.

In order to show that the molecular part of the sample really is in the $n=1$ state (with exactly one molecule at each lattice site), the molecules are first dissociated back into atom pairs by slowly (at 1.5~G/ms) ramping the magnetic field back across the Feshbach resonance. This brings the system back into the atomic Mott insulator state with shells with $n=1$ and $n=2$. Then, the atomic Mott insulator is melted by slowly (within 10~ms) ramping down the lattice from $V_0=24 E_r$ to $V_0=4 E_r$. Finally, the lattice is quickly switched off and after some time of flight an absorption image is taken.

\begin{figure}[h]
\includegraphics[height=3cm]{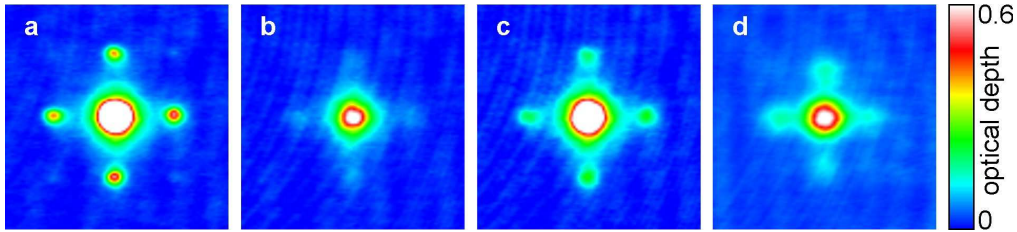}
\caption{\label{fig-restoration}
Restoration of phase coherence probing the quantum state with exactly one molecule per lattice site.
(a) An atomic Mott insulator is melted by reducing the lattice depth slowly. The system returns to the superfluid phase and phase coherence is restored. This phase coherence is probed by quickly switching the lattice off and observing an atomic interference pattern in time of flight.
(b) After association of molecules, only lattice sites occupied by $n=1$ atoms contribute to the signal.
(c) After association and dissociation of molecules, the satellite peaks are much stronger than in (b), thus proving that the molecular part of the cloud was in a molecular $n=1$ state.
(d) Pure molecular $n=1$ state. Same as (c) but between association and dissociation, remaining atoms were removed with blast light. Reproduced from Ref.~\cite{volz:06}.
 }
\end{figure}

\thispagestyle{plain}

Such images are shown in Fig.~\ref{fig-restoration}. Part (a) shows the result if the magnetic-field ramp for association and dissociation of molecules is omitted. This matter-wave interference pattern shows that phase coherence is restored when ramping down the lattice, thus demonstrating that an atomic Mott insulator is realized at 1008.8~G. Part (b) shows the pattern obtained if molecules are associated but not dissociated, so that they remain invisible in the detection. This signal comes only from sites with $n=1$ atoms. Part (c) shows the result obtained for the full sequence with association and dissociation of molecules. Obviously, the satellite peaks regain considerable population compared to (b), which proofs that after dissociation, we recover an atomic Mott insulator. This shows that association and dissociation must have been coherent and adiabatic. Combined with the freezing of the molecules' positions and the fact that the association starts from an atomic Mott insulator with an $n=2$ core, this implies that the molecular part of the cloud must have been in a quantum state with one molecule per lattice site.

After associating the molecules, remaining atoms can be removed from the trap using microwave radiation and a blast laser as in Refs.~\cite{xu:03,thalhammer:06}. This produces a pure molecular sample. The molecule numbers before and after the blast are identical within the experimental uncertainty of 5\%. In order to show that the pure molecular sample is in the $n=1$ state, the molecules are dissociated, the lattice is ramped down to $V_0=1.2 E_r$ within 30~ms, ramped back up to $V_0=6 E_r$ within 5~ms, and finally switched off. To understand why the extra ramping is needed, see Ref.~\cite{volz:06}. The result is shown in Fig.~\ref{fig-restoration}(d). Again, an interference pattern is visible. The time of flight was 12~ms in (a)-(c) and 11~ms in (d). The lifetime of the molecule number is 160(20)~ms, whereas the lifetime of the visibility of the interference pattern is 93(22)~ms \cite{volz:06}.

\begin{figure}[t]
\includegraphics[height=8.5cm]{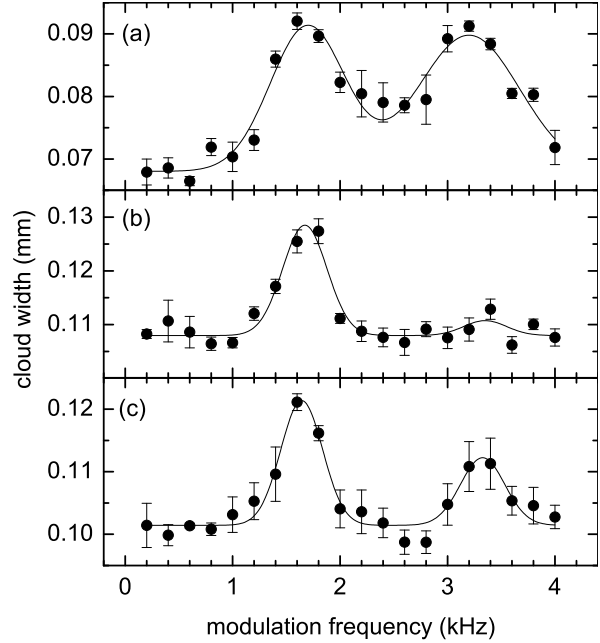}
\caption{\label{fig-excitation-atom}
Excitation spectrum of the atomic Mott insulator. The full width at half maximum of the central interference peak is shown as a function of the frequency at which the lattice depth is modulated. The results in parts (a)-(c) correspond to the conditions of Fig.~\ref{fig-restoration}(a)-(c). First, the usual lattice ramp-down starting at $V_0=24 E_r$ is interrupted at $V_0=15 E_r$. Next, the power of one lattice beam is modulated for 11~ms with a peak-to-peak amplitude of 50\%. Finally, the lattice ramp-down continues as usual. Resonances are visible at 1.6 and 3.2~kHz. The lines are a guide to the eye. Reproduced from Ref.~\cite{volz:06}.
}
\end{figure}

\thispagestyle{plain}

Figure~\ref{fig-excitation-atom} shows the excitation spectrum of the atomic Mott insulator at $V_0=15 E_r$ as measured by amplitude modulation of one lattice beam \cite{stoeferle:04}. To understand, how this spectroscopy method works, consider two neighboring lattice sites each initially occupied by exactly one atom. Without the modulation, the tunneling of one atom to the other site is suppressed because of the on-site interaction matrix element $U$, which shifts the tunneling process out of resonance. With the modulation, the lattice light has sidebands in frequency space. This makes carrier-sideband transitions possible, e.g., an absorption of a carrier photon by an atom can be followed by an induced emission into the red sideband. This can shift the above-described tunneling process into resonance, if the proper modulation frequency is chosen.

The spectrum in Fig.~\ref{fig-excitation-atom}(a) shows clear resonances at energies $U$ and $2U$. Below the first resonance, no noticeable excitations are created, showing that the excitation spectrum has a gap. This again shows that the system before molecule association is an atomic Mott insulator. In Fig.~\ref{fig-excitation-atom}(b), the signal at $2U$ essentially disappeared, because the signal at $2U$ is created by processes that require lattice sites with $n\geq2$ atoms \cite{greiner:02}, which are converted into molecules and hence absent. The spectrum in Fig.~\ref{fig-excitation-atom}(c) is similar to the one in Fig.~\ref{fig-excitation-atom}(a). Resonances at $U$ and $2U$ are clearly visible in Fig.~\ref{fig-excitation-atom}(c). This gives further experimental support for the above conclusion that the system after the association-dissociation ramp is still an atomic Mott insulator.

\begin{figure}[t]
\includegraphics[height=6.5cm]{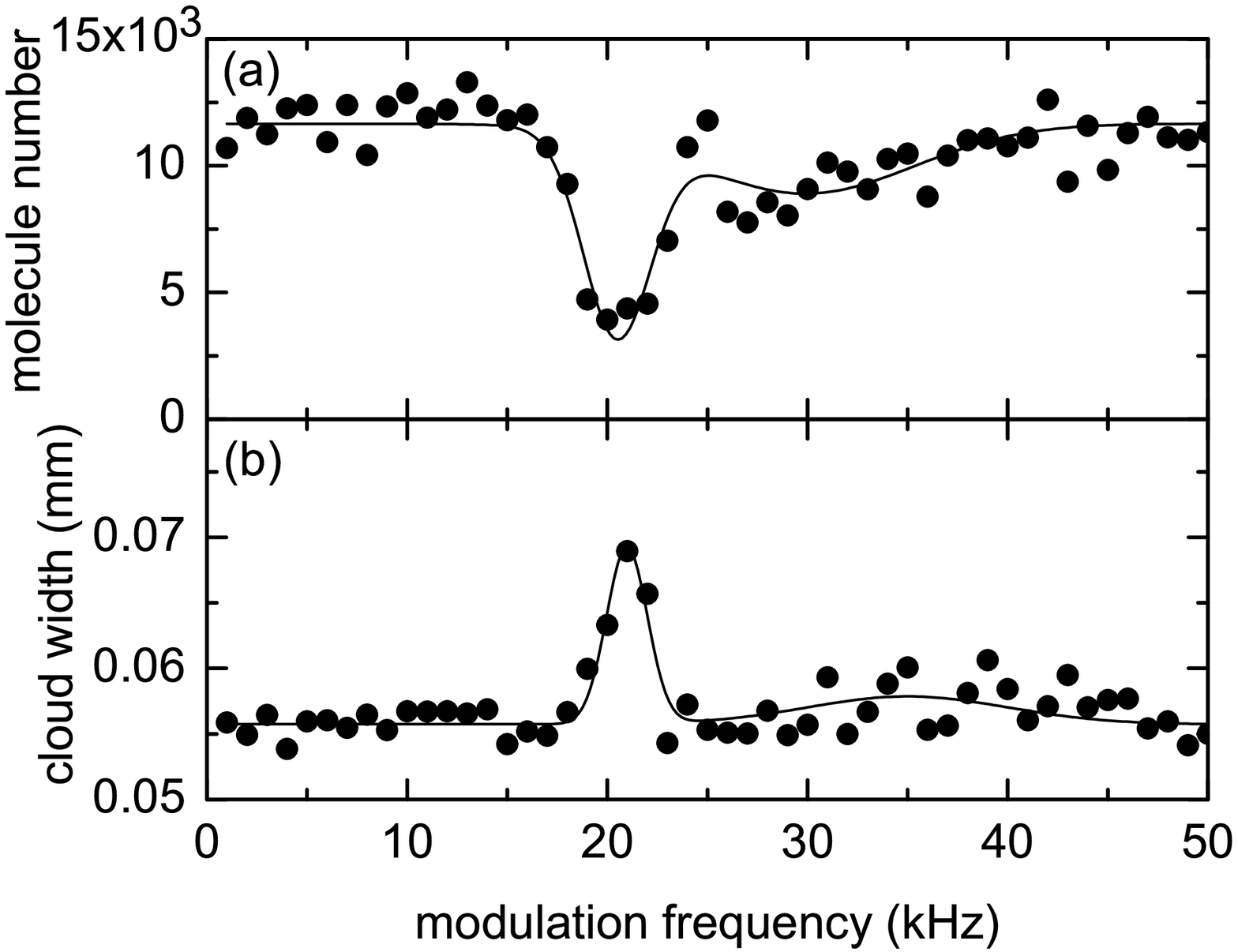}
\includegraphics[height=6.5cm]{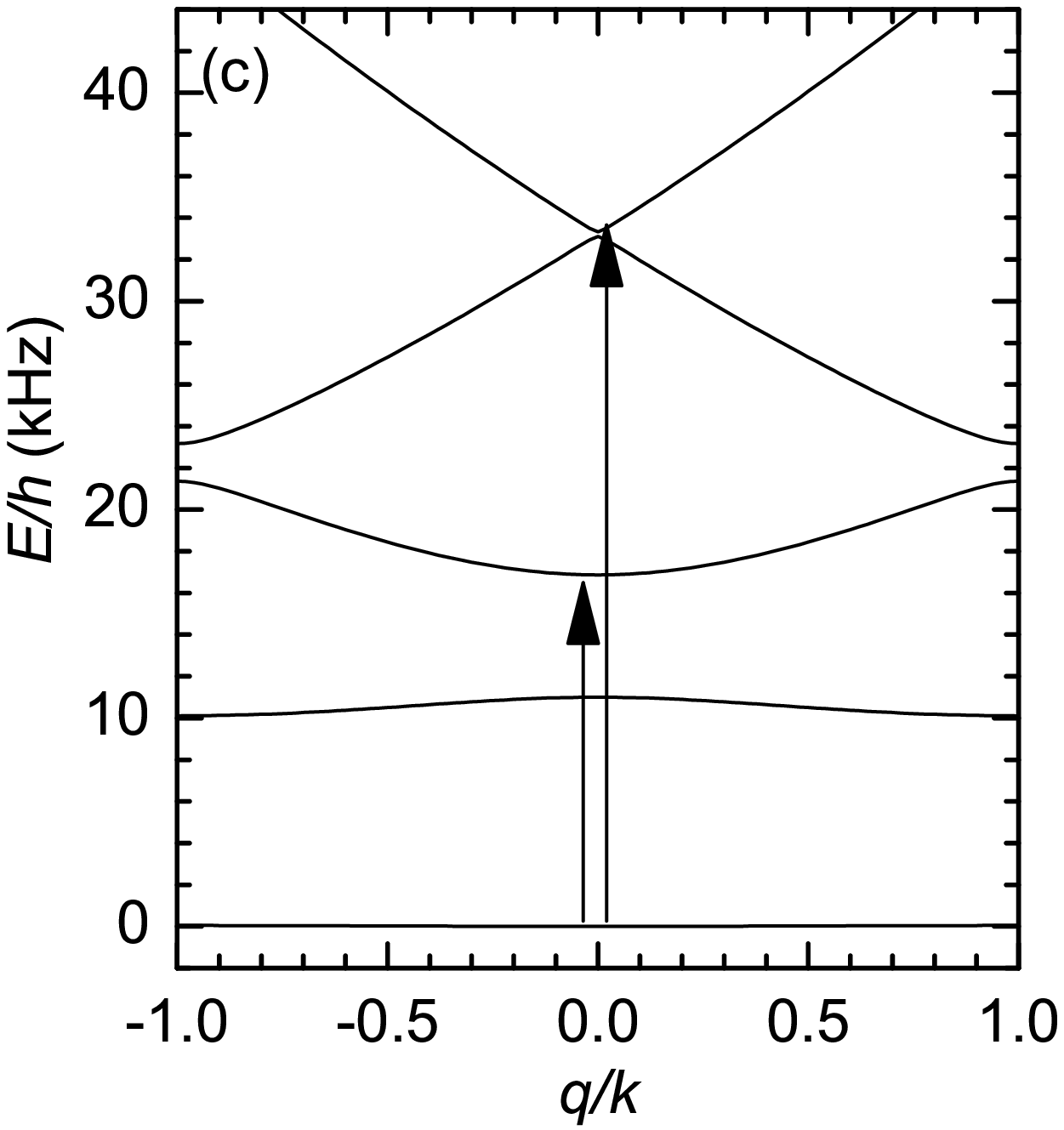}
\caption{\label{fig-excitation-molecule}
Excitation spectrum of the pure molecular $n=1$ state. Parameters are chosen similar to Fig.~\ref{fig-restoration}(d). The lattice is ramped down from $V_0=24 E_r$ to $V_0=3.6 E_r$ within 20~ms. Next, the power of one lattice beam is modulated for 10~ms with a peak-to-peak amplitude of 20\%. Then the lattice is switched off. Parts (a) and (b) show a narrow resonance at 21~kHz and a broad resonance around 33~kHz. The lines are a guide to the eye. These resonances are not related to the molecular on-site interaction matrix element $U_m$. Instead, both resonances can be explained as band excitation, when considering the one-dimensional band structure for molecules at this lattice depth, which is shown in part (c). The zero of energy in (c) is chosen at the bottom of the lowest band. Arrows indicate the band excitations.
 }
\end{figure}

\thispagestyle{plain}

We also measured excitation spectra at various lattice depths for the molecular $n=1$ state after removing the atoms, corresponding to Fig.~\ref{fig-restoration}(d). These spectra show no resonances related to the molecular on-site interaction matrix element $U_m$. An example is shown in Fig.~\ref{fig-excitation-molecule}. The lattice depth chosen here is a factor of approximately 4 lower than in Fig.~\ref{fig-excitation-atom}. Hence, the tunneling amplitude for the molecules $J_m=2\pi\hbar\times12$~Hz in Fig.~\ref{fig-excitation-molecule} is comparable to the tunneling amplitude for atoms $J=2\pi\hbar\times21$~Hz in Fig.~\ref{fig-excitation-atom}. The search for resonances related to $U_m$ is hampered by the fact that the value of $U_m$ is unknown. Furthermore, the lifetime of two molecules at one lattice site is very short \cite{volz:06}. This leads to an estimated resonance width of $\Gamma=2\pi\hbar \times 10$~kHz. Hence, a possible resonance would be too broad and consequently too shallow to be observed.

We took data in a broad parameter range, but could not find any resonances related to $U_m$. We only observed band-excitation resonances, as in Fig.~\ref{fig-excitation-molecule}. Both resonances in Fig.~\ref{fig-excitation-molecule} can be explained as band excitation. We took data in deeper lattices. Here, the broad resonance splits into two narrower resonances, which both follow the energy of a band excitation as a function of lattice depth. The molecule loss at the band-excitation resonance is caused by the fact that the tunneling amplitude in higher bands is much higher than in the lowest band. Hence, a molecule in a higher band can easily tunnel to a neighboring site. If this site was already occupied by another molecule, then an inelastic collision will lead to fast loss.

The data in Fig.~\ref{fig-excitation-molecule} do not show any signature of the first band-excitation resonance at 11~kHz. This is because of the parity of the states involved. To illustrated this, we consider the states with quasi-momentum $q=0$. These states have even/odd parity for even/odd band number. Therefore transitions between even and odd bands are suppressed.

In conclusion, we prepared a quantum state with exactly one molecule at each site of an optical lattice. The quantum state prepared here is exactly the state that a molecular Mott insulator has in the limit of negligible tunneling. Unlike the creation of a molecular Mott state by a quantum phase transition from a molecular BEC, our method works independently of the molecule-molecule interaction properties.


\thispagestyle{plain}


\begin{thebibliography}{29}
\expandafter\ifx\csname natexlab\endcsname\relax\def\natexlab#1{#1}\fi
\providecommand{\enquote}[1]{``#1''}
\expandafter\ifx\csname url\endcsname\relax
  \def\url#1{\texttt{#1}}\fi
\expandafter\ifx\csname urlprefix\endcsname\relax\def\urlprefix{URL }\fi
\providecommand{\eprint}[2][]{\url{#2}}

\bibitem[DeMille(2002)]{demille:02}
D.~DeMille, \emph{Phys. Rev. Lett.} \textbf{88}, 067901 (2002).

\bibitem[G{\'o}ral et~al.(2002)]{goral:02}
K.~G{\'o}ral, L.~Santos, and M.~Lewenstein, \emph{Phys. Rev. Lett.}
  \textbf{88}, 170406 (2002).

\bibitem[Lee and Ostrovskaya(2005)]{lee:05}
C.~Lee, and E.~A. Ostrovskaya, \emph{Phys. Rev. A} \textbf{72}, 062321 (2005).

\bibitem[Micheli et~al.(2006)]{micheli:06}
A.~Micheli, G.~K. Brennen, and P.~Zoller, \emph{Nature Phys.} \textbf{2},
  341--347 (2006).

\bibitem[Barnett et~al.(2006)]{barnett:06}
R.~Barnett, D.~Petrov, M.~Lukin, and E.~Demler, \emph{Phys. Rev. Lett.}
  \textbf{96}, 190401 (2006).

\bibitem[Regal et~al.(2003)]{regal:03}
C.~A. Regal, C.~Ticknor, J.~L. Bohn, and D.~S. Jin, \emph{Nature} \textbf{424},
  47--50 (2003).

\bibitem[Herbig et~al.(2003)]{herbig:03}
J.~Herbig, T.~Kraemer, M.~Mark, T.~Weber, C.~Chin, H.-C. N{\"a}gerl, and
  R.~Grimm, \emph{Science} \textbf{301}, 1510--1513 (2003).

\bibitem[D{\"u}rr et~al.(2004{\natexlab{a}})]{duerr:04}
S.~D{\"u}rr, T.~Volz, A.~Marte, and G.~Rempe, \emph{Phys. Rev. Lett.}
  \textbf{92}, 020406 (2004{\natexlab{a}}).

\bibitem[Strecker et~al.(2003)]{strecker:03}
K.~E. Strecker, G.~B. Partridge, and R.~G. Hulet, \emph{Phys. Rev. Lett.}
  \textbf{91}, 080406 (2003).

\bibitem[Cubizolles et~al.(2003)]{cubizolles:03}
J.~Cubizolles, T.~Bourdel, S.~J. Kokkelmans, G.~V. Shlyapnikov, and C.~Salomon,
  \emph{Phys. Rev. Lett.} \textbf{91}, 240401 (2003).

\bibitem[Jochim et~al.(2003)]{jochim:03a}
S.~Jochim, M.~Bartenstein, A.~Altmeyer, G.~Hendl, C.~Chin, J.~Hecker-Denschlag,
  and R.~Grimm, \emph{Phys. Rev. Lett.} \textbf{91}, 240402 (2003).

\bibitem[Xu et~al.(2003)]{xu:03}
K.~Xu, T.~Mukaiyama, J.~R. Abo-Shaeer, J.~K. Chin, D.~E. Miller, and
  W.~Ketterle, \emph{Phys. Rev. Lett.} \textbf{91}, 210402 (2003).

\bibitem[Kohler et~al.(2006)]{koehler:06}
T.~Kohler, K.~Goral, and P.~S. Julienne, \emph{Rev. Mod. Phys.} \textbf{78},
  1311 (2006).

\bibitem[Sage et~al.(2005)]{sage:05}
J.~M. Sage, S.~Sainis, T.~Bergeman, and D.~DeMille, \emph{Phys. Rev. Lett.}
  \textbf{94}, 203001 (2005).

\bibitem[Greiner et~al.(2002)]{greiner:02}
M.~Greiner, O.~Mandel, T.~Esslinger, T.~W. H{\"a}nsch, and I.~Bloch,
  \emph{Nature} \textbf{415}, 39--44 (2002).

\bibitem[Thalhammer et~al.(2006)]{thalhammer:06}
G.~Thalhammer, K.~Winkler, F.~Lang, S.~Schmid, R.~Grimm, and
  J.~{Hec\-ker-Denschlag}, \emph{Phys. Rev. Lett.} \textbf{96}, 050402 (2006).

\bibitem[Rom et~al.(2004)]{rom:04}
T.~Rom, T.~Best, O.~Mandel, A.~Widera, M.~Greiner, T.~W. H{\"a}nsch, and
  I.~Bloch, \emph{Phys. Rev. Lett.} \textbf{93}, 073002 (2004).

\bibitem[St{\"o}ferle et~al.(2006)]{stoferle:06}
T.~St{\"o}ferle, H.~Moritz, K.~G{\"u}nter, M.~K{\"o}hl, and T.~Esslinger,
  \emph{Phys. Rev. Lett.} \textbf{96}, 030401 (2006).

\bibitem[Winkler et~al.(2006)]{winkler:06}
K.~Winkler, G.~Thalhammer, F.~Lang, R.~Grimm, J.~{Hecker Denschlag}, A.~J.
  Daley, A.~Kantian, H.~P. B{\"u}chler, and P.~Zoller, \emph{Nature}
  \textbf{441}, 853--856 (2006).

\bibitem[Ryu et~al.(2005)]{ryu:cond-mat/0508201-aipproc}
C.~Ryu, X.~Du, E.~Yesilada, A.~M. Dudarev, S.~Wan, Q.~Niu, and D.~J. Heinzen,
  \emph{{\rm Preprint at
  $\langle$http://arXiv.org/abs/cond-mat/0508201$\rangle$}}  (2005).

\bibitem[Jaksch et~al.(2002)]{jaksch:02}
D.~Jaksch, V.~Venturi, J.~I. Cirac, C.~J. Williams, and P.~Zoller, \emph{Phys.
  Rev. Lett.} \textbf{89}, 040402 (2002).

\bibitem[Damski et~al.(2003)]{damski:03}
B.~Damski, L.~Santos, E.~Tiemann, M.~Lewenstein, S.~Kotochigova, P.~Julienne,
  and P.~Zoller, \emph{Phys. Rev. Lett.} \textbf{90}, 110401 (2003).

\bibitem[Volz et~al.(2006)]{volz:06}
T.~Volz, N.~Syassen, D.~M. Bauer, E.~Hansis, S.~D{\"u}rr, and G.~Rempe,
  \emph{Nature Phys.} \textbf{2}, 692--695 (2006).

\bibitem[Marte et~al.(2002)]{marte:02}
A.~Marte, T.~Volz, J.~Schuster, S.~D{\"u}rr, G.~Rempe, E.~G. van Kempen, and
  B.~J. Verhaar, \emph{Phys. Rev. Lett.} \textbf{89}, 283202 (2002).

\bibitem[Volz et~al.(2003)]{volz:03}
T.~Volz, S.~D{\"u}rr, S.~Ernst, A.~Marte, and G.~Rempe, \emph{Phys. Rev. A}
  \textbf{68}, 010702(R) (2003).

\bibitem[D{\"u}rr et~al.(2004{\natexlab{b}})]{duerr:04a}
S.~D{\"u}rr, T.~Volz, and G.~Rempe, \emph{Phys. Rev. A} \textbf{70}, 031601(R)
  (2004{\natexlab{b}}).

\bibitem[Jaksch et~al.(1998)]{jaksch:98}
D.~Jaksch, C.~Bruder, J.~I. Cirac, C.~W. Gardiner, and P.~Zoller, \emph{Phys.
  Rev. Lett.} \textbf{81}, 3108--3111 (1998).

\bibitem[{Hecker-Denschlag} et~al.(2002)]{hecker:02}
J.~{Hecker-Denschlag}, J.~E. Simsarian, H.~H{\"a}ffner, C.~McKenzie,
  A.~Browaeys, D.~Cho, K.~Helmerson, S.~L. Rolston, and W.~D. Phillips,
  \emph{J. Phys. B} \textbf{35}, 3095--3110 (2002).

\bibitem[St{\"o}ferle et~al.(2004)]{stoeferle:04}
T.~St{\"o}ferle, H.~Moritz, C.~Schori, M.~K{\"o}hl, and T.~Esslinger,
  \emph{Phys. Rev. Lett.} \textbf{92}, 130403 (2004).

\end{thebibliography}

\end{document}